%% file: su6DM_rev.tex
\documentclass[reprint,twocolumn,superscriptaddress,showpacs,amsmath,amssymb,aps,prl]{revtex4-1}

\usepackage{bbm}

\usepackage{graphicx,epstopdf}
\usepackage{epsfig}
\usepackage{subfigure}
\usepackage[usenames,dvipsnames]{color}
\usepackage{slashed}
\usepackage{hyperref}
\usepackage[utf8]{inputenc}
\usepackage{multirow}

\allowdisplaybreaks[2]

\newcommand{\beq}{\begin{eqnarray}}
\newcommand{\eeq}{\end{eqnarray}}

\newcommand{\bmp}{\noindent\begin{minipage}{16cm}}
\newcommand{\emp}{\end{minipage}\vskip 7mm} 

\usepackage{dcolumn}
\usepackage{bm}
\usepackage[normalem]{ulem}

\begin{document}

\title{ A Singlet  Dark Matter in the SU(6)/SO(6) Composite Higgs Model}

\author{Haiying Cai}
\email{haiying.cai@apctp.org}
\affiliation{Asia Pacific Center for Theoretical Physics, Pohang, Gyeongbuk 790-784, Republic of Korea}
\affiliation{Tsung-Dao Lee Institute, and School of Physics and Astronomy, Shanghai Jiao Tong University, Shanghai 200240, China}

\author{Giacomo Cacciapaglia}
\email{g.cacciapaglia@ipnl.in2p3.fr}
\affiliation{Institut de Physique des Deux Infinis de Lyon (IP2I), CNRS/IN2P3 UMR5822, 4 rue Enrico Fermi,
69622 Villeurbanne Cedex, France} 

\affiliation{Universit\'e Claude Bernard Lyon 1, Universit\'e de Lyon, 92 rue Pasteur, 69361 Lyon Cedex 07, France}

\begin{abstract}
{ 
Singlet scalar Dark Matter can naturally arise in composite Higgs models as an additional stable pseudo-Nambu-Goldstone boson. We study the properties of such a candidate in a model based on $SU(6)/SO(6)$, with the light quark masses generated by 4-fermion interactions. The presence of non-linearities in the couplings allows to saturate the relic density for masses $400 < m_{\rm DM} < 1000$~GeV, and survive the bound from Direct Detection and Indirect Detection. The viable parameter regions are in reach of the sensitivities of future upgrades, like XENONnT and LZ.
}
\end{abstract}


\maketitle

\section{Introduction}

The standard cosmology model, ``$\Lambda$CDM''  based on a flat prior, can well describe an expanding universe from the early to late times.  The combination of radiation, matter and dark energy determines the Hubble expansion  $H(t) = \dot a(t) / a(t)$  as governed by  the Friedmann equations. According to  astrophysical measurements, the present Universe  consists of roughly $30\%$ matter  and  $70\%$  dark energy after the dark age and large scale structure emergence~\cite{Aghanim:2018eyx}.   However, the density of baryonic matter today is $\Omega_b h^2 \sim 0.02$, comprising only a small portion of the total matter. Thus, the remaining $85\%$ of the total matter is made of a Dark Component,  expected to be distributed as spherical halos around Galaxies. Despite the convincing  evidences for  Dark Matter (DM) from various sources, such as  the galaxy rotation curves, gravitational lensing and  observations of cosmic microwave background (CMB), the particle identity of DM has not been identified yet. 
In the Standard Model of Particle Physics (SM), the only candidates, neutrinos, are too light and have too small a relic density to account for the observation. Therefore, many theories Beyond the Standard Model (BSM) have been proposed, with the most popular ones advocating weakly interacting massive particles (WIMPs) stabilised by a discrete symmetry: this scenario can naturally provide a suitable DM candidate thanks to the thermal decoupling. 

One of the  possibilities is the Higgs mediated singlet Scalar DM model. Without consideration for the naturalness of the light scalar mass, the DM physics is simply described by two free parameters: the mass $m_{S}$ and Higgs coupling to the singlet scalar $\lambda_{hSS}$~\cite{Silveira:1985rk,McDonald:1993ex,Burgess:2000yq}. Because of the small parameter space, this model has high predictive power. The dominant DM annihilation channels, which determine the thermal relic density, are as follows:
for  $m_S < m_W $, a DM pair mainly annihilates into $\bar b b$, while in the high mass region its annihilation cross section into $WW$, $ZZ$ and $hh$  turns to be very effective.  However,  the viable parameter space for the model is tightly squeezed~\cite{Casas:2017jjg, Arcadi:2019lka} since the  $\lambda_{hSS}$ coupling is subject to strong constraints from  direct detection experiments, e.g. Xenon1T~\cite{Aprile:2018dbl}, PandaX~\cite{Cui:2017nnn} and LUX~\cite{Akerib:2016vxi}, as well as by the bounds on the Higgs invisible decay width and the upper bounds on events with large missing energy at the LHC experiments. In this work  we plan to investigate this scenario in the context of a Composite Higgs Model (CHM) that enjoys an underlying gauge-fermion description and can be UV completed. The scalar DM candidate emerges as a pseudo-Nambu-Goldstone boson (pNGB) together with the Higgs boson itself.  In our scenario, the composite nature of both DM and the Higgs boson can substantially modify  the DM couplings to the SM states and alter the relative importance of various annihilation channels. This is mainly due to the presence of higher order couplings, generated by non linearities in the pNGB couplings, which can enhance the annihilation cross sections while the coupling to the Higgs (constrained by Direct Detection) is small. This kind of scenarios was first proposed in Ref.~\cite{Frigerio:2012uc} in the context of the minimal coset $SU(4)/Sp(4)$, however the scalar candidate is allowed to decay if a topological anomaly is present, like it is always the case for models with a microscopic gauge-fermion description~\cite{Galloway:2010bp}. Thus, it is necessary to work in scenarios with larger global symmetries, which allows for singlet scalar states which do not couple via the topological term, and can therefore be stable.  Also, in Ref.~\cite{Balkin:2018tma} it has been pointed out that the DM has dominant derivative couplings to the Higgs, which again ensures the suppression of Direct Detection rates~\cite{Gross:2017dan}. However, the nature of the couplings is basis dependent, as one can always choose a basis for the pNGBs where the derivative coupling is absent~\cite{Cacciapaglia:2020kgq}. In this paper we will work in this basis.

The CHM we studied is based on the coset $SU(6)/SO(6)$, which enjoys a microscopic description with underlying fermions transforming in a real representation of the confining gauge group~\cite{Cacciapaglia:2019ixa}.  Top partial compositeness can also be implemented along the lines of~\cite{Ferretti:2013kya}.
The pNGB sector is similar to the $SU(5)/SO(5)$ model~\cite{Dugan:1984hq}, which is the minimal realistic coset in the $SU(N)/SO(N)$ family: the pNGBs include a bi-triplet of the custodial $SU(2)_L\times SU(2)_R$ global symmetry, like the Georgi-Machacek (GM) model~\cite{Georgi:1985nv}. However, unlike the GM model, the direction inside the bi-triplet that does not violate the custodial symmetry is CP-odd, thus it usually cannot develop a vacuum expectation value in a CP conserving theory. On the other hand, the interactions of fermions to the composite sector typically induce a tadpole for the custodial triplet component, thus generating unbearable contributions to the $\rho$ parameter. For the top, coupling to composite fermions in the adjoint representation of $SU(5)$ allows to avoid this issue~\cite{Agugliaro:2018vsu}. 
In our model, the adjoint of $SU(6)$ serves the same purpose, while the masses of the light fermions can be generated by other mechanisms. This results in a violation of custodial symmetry of the order of $m_b^2/m_h^2$, thus being small enough to evade precision bounds, as we demonstrate in this paper. The extension of the model to $SU(6)$ also contains a second Higgs doublet and a singlet, which can be protected by a $\mathbb{Z}_2$ symmetry for suitable couplings of the top quark~\cite{Cacciapaglia:2019ixa}.
For other examples of CHMs with DM, see Refs~\cite{Marzocca:2014msa,Ma:2017vzm,Ballesteros:2017xeg,Balkin:2017aep, Balkin:2017yns, Alanne:2018zjm, Cai:2019cow, Ramos:2019qqa}. 

In this work, we investigate the properties of the singlet, $\mathbb{Z}_2$--odd, pNGB as candidate for Dark Matter. We find that, notwithstanding the presence of additional couplings, the model is tightly constrained, especially by direct detection. The small parameter space still available will be tested by the next generation direct detection experiments, with DM masses in the $400$ to $1000$ GeV range.

\section{The model}\label{model}

The main properties of the low energy Lagrangian associated to this model have been studied in detail in~\cite{Cacciapaglia:2019ixa},
where we refer the reader for more details. In this section, we will briefly recall the main properties of the pNGBs, and discuss in detail
how custodial violation is generated via the masses of the light SM fermions. The latter point was not discussed in the previous work.
Following Refs~\cite{Agugliaro:2018vsu,Cacciapaglia:2019ixa}, we will embed the SM top fields in the adjoint representation of the global
$SU(6)$: this is the only choice that allows for vanishing triplet VEV, thus preserving custodial symmetry. For the light fermions, we will
add direct four-fermion interactions, to generate effective Yukawa couplings to the composite Higgs sector.

For a start, we recall the structure of the 20 pNGBs generated in this model. To do so, it is convenient to define them around a vacuum
that preserves the EW symmetry, incarnated in a $6\times 6$ symmetric matrix $\Sigma_{\rm EW}$ (for the explicit form, see~\cite{Cacciapaglia:2019ixa}). The  pNGBs can thus be classified in terms of the EW gauge symmetry $SU(2)_L \times U(1)_Y$, and of the global custodial symmetry 
envelope $SU(2)_L \times SU(2)_R$, which needs to be present in order to preserve the SM relation between the $Z$ and $W$ masses~\cite{Georgi:1984af}. These quantum numbers are given in Table~\ref{tab:pNGBs}.

\begin{table}[tbh]
\begin{tabular}{c|c|c|c|}
 & $SU(2)_L \times U(1)_Y$ & $SU(2)_L \times SU(2)_R$ & $\mathbb{Z}_2$ \\ \hline \hline
$H_1$ & $(2,\pm 1/2)$ & $(2,2)$ & $+$ \\ \hline
$H_2$ & $(2,\pm 1/2)$ & $(2,2)$ & $-$ \\ \hline
$\Lambda$ & $(3,\pm 1)$ & \multirow{2}{*}{$(3,3)$} & \multirow{2}{*}{$+$} \\ 
$\varphi$ & $(3, 0)$ &  & \\ \hline
$\eta_1$ & $(1,0)$ & $(1,1)$ & $+$ \\ \hline
$\eta_2$ & $(1,0)$ & $(1,1)$ & $-$ \\ \hline
$\eta_3$ & $(1,0)$ & $(1,1)$ & $+$ \\ \hline
\end{tabular}
\caption{Quantum numbers of the 20 pNGBs in terms of the EW and custodial symmetries. Here, the pNGB are defined around the EW-preserving vacuum $\Sigma_{\rm EW}$. The last column indicates the $\mathbb{Z}_2$ protecting the DM candidate, as defined in the text.} \label{tab:pNGBs}
\end{table}

A non-linearly transforming pNGB matrix can be defined as
\begin{equation}
\Sigma (x) = e^{i\frac{2 \sqrt{2}}{f} \Pi(x)} \cdot \Sigma_{\rm EW}\,,
\end{equation}
where $\Pi(x)$ contains the pNGB fields~\cite{Cacciapaglia:2019ixa}. We can now define a $\mathbb{Z}_2$ transformation that is from a broken global $U(1)$:
\begin{eqnarray}
\Omega_{\rm DM} = \left( \begin{array}{ccc}
\mathbbm{1}_{2} &   &  \\
& \mathbbm{1}_{2} &   \\
&   & \sigma_{3}  \\
\end{array} \right), \;\;  \Omega_{\rm DM} \Sigma(x)\Omega_{\rm DM} =  \Sigma' (x)\,, \label{eq: parity}
\end{eqnarray}
where $H_2 \to - H_2$ and $\eta_2 \to -\eta_2$ in $\Sigma' (x)$. Thus, they are the $\mathbb{Z}_2$--odd states, while all the other pNGBs are even, as indicated in the last column in Table~\ref{tab:pNGBs}.
This parity commutes with the EW and custodial symmetries, and with a suitable choice of the top couplings in the adjoint spurion~\cite{Cacciapaglia:2019ixa}, thus it can remain an exact symmetry of this model. This $\mathbb{Z}_2$  with $\det \Omega_{\rm DM} = -1$ is a remnant of a  $U(1)$ global symmetry, that  protects  DM candidates from  the topological anomaly interaction in the microscopic gauge-fermion theory and uniquely determines the parity  assignment in Table~\ref{tab:pNGBs}. The DM candidates, therefore, can be either the singlet pseudo-scalar $\eta_2$ or the component field $A_0$ in the second doublet $H_2$,  which are both CP-odd.  Note that the CP-even components of  $H_{\pm, 0}$ are always heavier and  will eventually decay into the lightest $\mathbb{Z}_2$-odd particle.

To study the properties of the DM candidate, we need to introduce the effects due to the breaking of the EW symmetry. The latter is due to some pNGBs acquiring a VEV: this effects can be introduced as a rotation of the vacuum by a suitable number of angles. In our case, as we want to preserve the Dark $\mathbb{Z}_2$, we will assume that only $H_1$ will acquire a VEV, and check the consistency of this choice {\it a posteriori} when studying the potential for the pNGBs. For generality, we also introduce a VEV for the custodial triplet, corresponding to $\lambda_0 = - \frac{i}{\sqrt{2}}(\Lambda_0 - \Lambda_0^*)$.
The rotation allows to define a new pNGB matrix
\begin{equation}
\Sigma_{\alpha,\gamma} (x) = U (\alpha, \gamma) \cdot \Sigma (x) \cdot U^T (\alpha, \gamma)\,,
\end{equation}
where
\begin{equation} \label{eq:U}
U(\alpha,\gamma) =  e^{i \gamma \sqrt{2} S_{8}}\cdot e^{i \alpha \sqrt{2} X_{10}}\cdot e^{-i \gamma \sqrt{2} S_{8}}\,.
 \end{equation}
This rotation can be interpreted as follows: the exponential containing $\alpha$ is generated by a VEV for $H_1$, aligned with the broken generator $X_{10}$; then, we operate a rotation generated by the unbroken generator $S_8$ that misaligns the VEV along the direction of $\lambda_0$ by an angle $\gamma$.  As the effect of the VEVs is an $SU(6)$ rotation of the pNGB matrix, the couplings among pNGBs are unaffected as the chiral Lagrangian is invariant under such type of rotation: thus, no derivative couplings between one Higgs boson with two DM states is generated in this basis. Here we normalise $f$ such that $v= 246~\mbox{GeV}= f \sin \alpha$ for $\gamma = 0$.

The precise value of the two angles is determined by the total potential, generated by couplings that explicitly break the global symmetry $SU(6)$. After turning on the $\gamma$,  we can evaluate the potential in a new basis $\tilde{\Pi}(x)$, which is equivalent to  a pNGB field redefinition,
\begin{eqnarray}
\tilde{h} &=& h \cos \gamma - \lambda_0 \sin \gamma\,,  \nonumber \\ 
\tilde{\lambda}_0 &=&  h \sin \gamma + \lambda_0 \cos \gamma\,,
\end{eqnarray}
with the full mapping of other pNGBs  provided in the Appendix.  We proved that, as expected, only tadpoles for $\tilde{h}$ (i.e.  the  Higgs) and $\tilde{\lambda}_0$ are generated. Furthermore, the tadpole terms from  all  contributions observe the same structure:  $\frac{1}{f} \frac{\partial V_0(\alpha, \gamma)}{\partial \alpha} \tilde{h} - \frac{1}{v} \frac{\partial V_0(\alpha, \gamma)}{\partial \gamma} \tilde{\lambda}_0$, like the Taylor expansions.  Thus, the vanishing of the tadpoles is guaranteed at the minimum of the potential. This is a main result of this paper and  validates our choice for the vacuum misalignment in Eq.~\eqref{eq:U}.

First, we study the potential $V_0$ coming from the top, gauge and underlying fermion mass, at $\gamma=0$  as studied in~\cite{Cacciapaglia:2019ixa}. The $V_0$ depends on 4 independent parameters: $Q_A$ and $R_S$ in the top sector, the underlying mass $Bm$ (where $B$ is a dimension-less form factor while $m$ is the value of the underlying fermion mass), and a form factor for the gauge loops $C_g$. The $Q_A$ represents the coupling of the quark doublet to  top partner, and the $R_S$ is the coupling of the quark singlet (a second parameter $R'_S \equiv r R_S$  is irrelevant to $V_0$).  Although $C_g$ and $B$ can be computed on the lattice once an underlying dynamics is fixed,  we will  treat them as  free parameters. The $Q_A$ and $R_S$  can be traded in terms of the misalignment angle $\alpha$  and the Higgs boson mass $m_h$:
\begin{eqnarray}
Q_{A}^2 &=& \frac{m_{h}^2 }{6 v^2} \frac{\cos (2 \alpha )}{\cos^2(\alpha )}  -\frac{8 \sqrt{2} B  m  \sin (\alpha )  }{3 v}  \nonumber \\
&&  - \frac{ C_g g_{2}^2  \left(\cos
   \left(2 \theta _W\right)+2\right) }{3 \cos^2\left(\theta _W\right)}\,, \\
  R_{S}^2 &=& 4 ~ Q_{A}^2 + \frac{m_h^2 }{2 v^2}  \sec^2(\alpha ) \,,
\end{eqnarray}
where $m_h = 125$~GeV is the measured Higgs mass.

We now introduce the masses for the light fermions via direct couplings. This means, in practice, that we introduce effective Yukawa couplings: e.g., for the bottom quark
\begin{equation}
- Y_b\; (\bar{q}_L b_R)\ \mbox{Tr} [P \cdot \Sigma_{\alpha ,\gamma}(x)] + \mbox{h.c.}
\end{equation} 
where $Y_b$ is the Yukawa coupling and $P$ is a matrix in the $SU(6)$ space that extracts the Higgs components out of the pNGB matrix $\Sigma_{\alpha ,\gamma}(x)$~\cite{Cacciapaglia:2019ixa}.  Note that this kind of operators may also derive from partial compositeness upon integrating out the heavy partners of the light quarks. At one loop, this will generate a contribution to the potential for $\alpha$ and $\gamma$ of the form
\begin{equation}
V_b \approx f^4 C_b \left| \mbox{Tr} [P \cdot \Sigma_{\alpha, \gamma} (x)] \right|^2\,,
\end{equation}
which generates a tadpole for $\tilde{\lambda}_0$ that does not vanish for $\gamma=0$. More details on all potential contributions can be found in the Appendix.

Expanding for small $\alpha$ and small $\gamma$, the cancellation of the tadpole $\frac{\partial V_0(\alpha, \gamma)}{\partial \gamma}=0$ yields the following result
\begin{eqnarray}
\gamma \simeq \frac{12 \, \alpha \, m_{b}^2}{C_{g} m_{W}^2  (72 + 40 \tan ^2\left(\theta _W\right) )+ m_{h}^2} \,.
\end{eqnarray}
As a consequence of a non-vanishing $\gamma$, the model suffers from a tree-level correction to the $\rho$ parameter:
\begin{eqnarray}
\delta \rho_\gamma \simeq \frac{\sin ^2(\gamma )+1}{(2 \cos (2 \alpha )+1) \sin ^2(\gamma )+1}-1\,.
\end{eqnarray}
To study the impact of this correction, we determined the constraints  from the EW precision tests, in the form of the oblique parameters $S$ and $T$. Besides the tree level correction, which impact directly $T$, we also included loops deriving from the modification of the Higgs couplings to gauge bosons, and a generic contribution to $S$ from the strong sector. We thus define~\cite{Arbey:2015exa}:
\begin{eqnarray}
S &=& \frac{\sin^2 \alpha}{6 \pi} \left(\ln \frac{4 \pi f}{m_h} + N_D \right)\,, \\
T &=& - \frac{3\sin^2 \alpha}{8 \pi \cos^2 \theta_W} \ln \frac{4 \pi f}{m_h} + \frac{\delta \rho_\gamma}{\alpha_{\rm em}}  \,,
\end{eqnarray}
where $N_D$ counts the number of EW doublets in the underlying theory. 

\begin{figure}[tb]
	\centering 
	\includegraphics[height=6.3cm, 
		width=7.0cm]{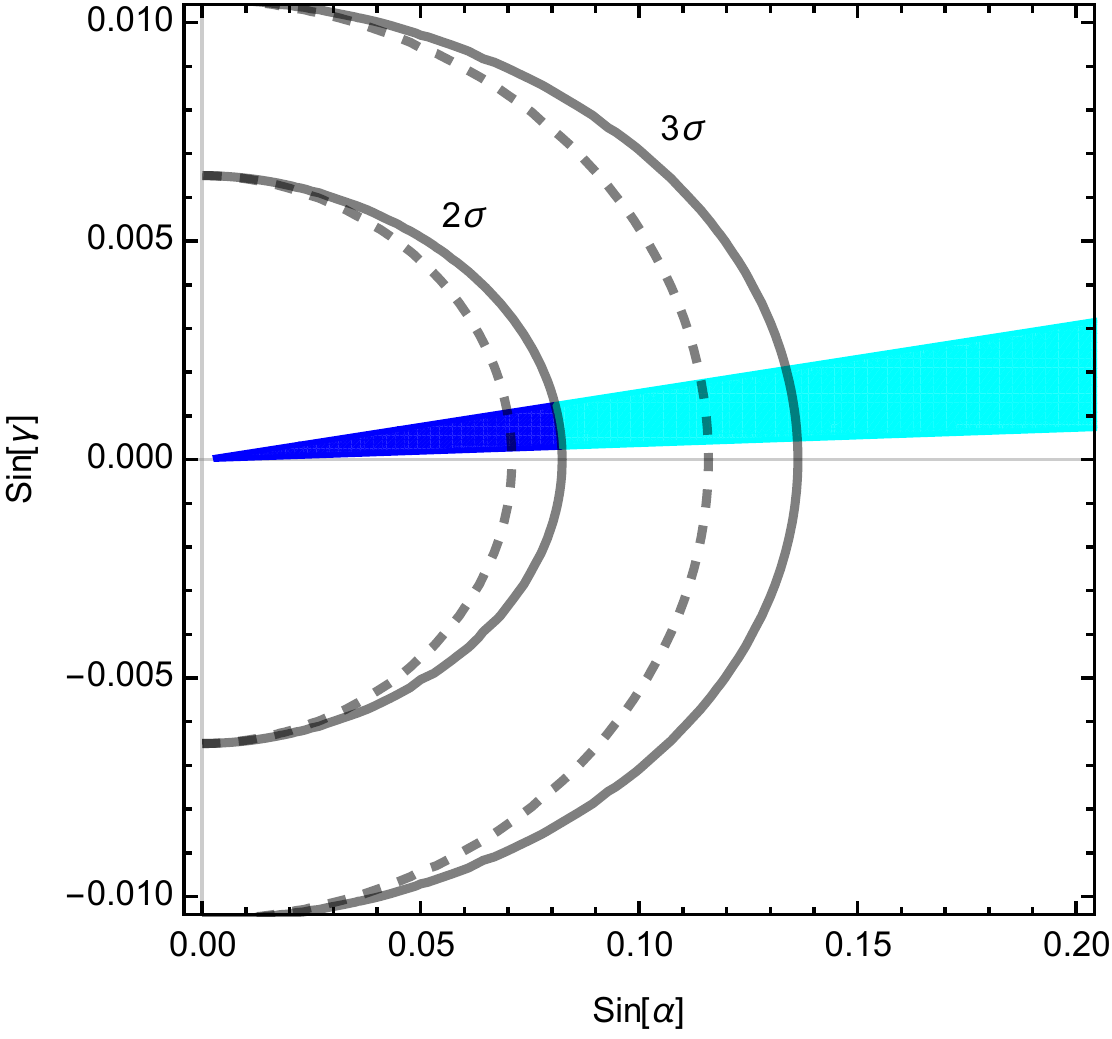}	
		\caption{EW precision bounds in the $\sin \alpha$--$\sin \gamma$ plane for model M3 (solid) and M4 (dashed). The blue wedge is the predicted region in the model, where the dark region is allowed at $2\sigma$. We vary $C_g$ between $0$ (top edge) to $0.1$ (bottom edge).} \label{fig:EWPTs}
\end{figure}

Numerically, we plot the bounds in Fig.~\ref{fig:EWPTs} for two realistic models~\cite{Ferretti:2013kya}: only two models are relevant, both based on a confining $\mathcal{G}_{\rm HC} = SO(N_c)$, with $N_c = 7,9$, and underlying fermions in the spinorial representation. Following the nomenclature of~\cite{Belyaev:2016ftv,Cacciapaglia:2019bqz}, we show M3 ($N_D = 16$) in solid and M4 ($N_D = 32$) in dashed. The blue shaded wedge is the region or parameters spanned in our model, where we vary $0 < C_g < 0.1$. The plot shows that the value of $\gamma$ is always very small, and that the bound on the parameter space is always dominated by the contribution of $\alpha$ to the $S$ parameter.  We should note that the generic contribution of the strong dynamics to the $S$ parameter can be reduced in various way: replacing it by loops of the heavy states~\cite{Ghosh:2015wiz}, considering a cancellation between vector and axial resonances~\cite{Hirn:2006nt}, or including the effect of a light-ish $0^{++}$ state~\cite{BuarqueFranzosi:2018eaj}.
For our purposes, the main point is to show that the effect of custodial breaking via $\gamma$ is under control. We will not consider the bound on $\alpha$ from EW precision in the following because it can be reduced in a model-dependent way.


\section{Dark Matter  phenomenology}\label{DMpheno}

Since the $\mathbb{Z}_2$--odd states in our model have sizable couplings to the SM, the relic abundance can be produced by the thermal freeze-out mechanism. 
We recall that the freeze-out temperature is typically a fraction of the DM mass, and that the DM mass in this model is always at most of the same order as the compositeness scale $f$. Thus, the calculations in this section can be performed within the range of validity of the effective theory, that is trustable up to $\Lambda_{\rm EFT} \sim 4 \pi f$.
The DM candidate(s) remain in thermal equilibrium with the SM, until the DM annihilation rate drops below the Hubble expansion rate. After the decoupling from the thermal bath (freeze-out), the DM density remains constant in a co-moving volume. Defining the yield $Y = n(x) /s (x)$,  where $s(x)$ is the entropy density,  the DM density evolution  is described by the Boltzmann equation~\cite{Kolb:1990vq}:
\begin{equation}
\frac{d Y}{d x}  = -  \frac{s(x=1)}{x^2 H(x=1)} \langle \sigma_{\rm eff} v \rangle [Y^2 - Y_{\rm{eq}}^2]\,, 
\end{equation}
where
\begin{equation}
s(x = 1) = \frac{2 \pi^2}{45} g_* M^3  \,, \;\; H (x = 1) = \sqrt{\frac{\pi^2}{90} g_*} \frac{M^2}{M_{\rm pl}}\,,
\end{equation}
with $x= M/T$ and $M$ being the DM mass. The effective thermal averaged cross-section can normally be expanded as $\langle \sigma_{\rm eff} v \rangle = a^{(0)} + \frac{3}{2} a^{(1)} x^{-1} + \cdots $, unless the particle masses are near a threshold or a resonant regime~\cite{Griest:1990kh}. Using an analytic approach, the relic density is calculated to be: 
\begin{align}
&\Omega h^2 = \frac{1.07\times 10^9 {\rm GeV}^{-1}}{\sqrt{g_*(x_f)} M_{pl} J (x_f)}\,,   \quad J (x_f)= \int_{x_f}^\infty dx \frac{\langle\sigma_{\rm eff} v \rangle}{x^2} \,.
\end{align}
Assuming s-wave dominance for the annihilation cross-sections, the above equations yield an approximate solution $ \langle \sigma_{\rm eff} v \rangle \simeq 2.0 \times 10^{-26} {\rm{cm}}^3/s$  in order to saturate  the relic density $\Omega h^2 = 0.12 \pm 0.001$ observed in the present universe \cite{Aghanim:2018eyx}.
In the following we will use these results to calculate the favourable region of parameter space in our model.

\begin{figure*}[tb]
	\centering 
	\includegraphics[height=5.8 cm, 
		width=7.0cm]{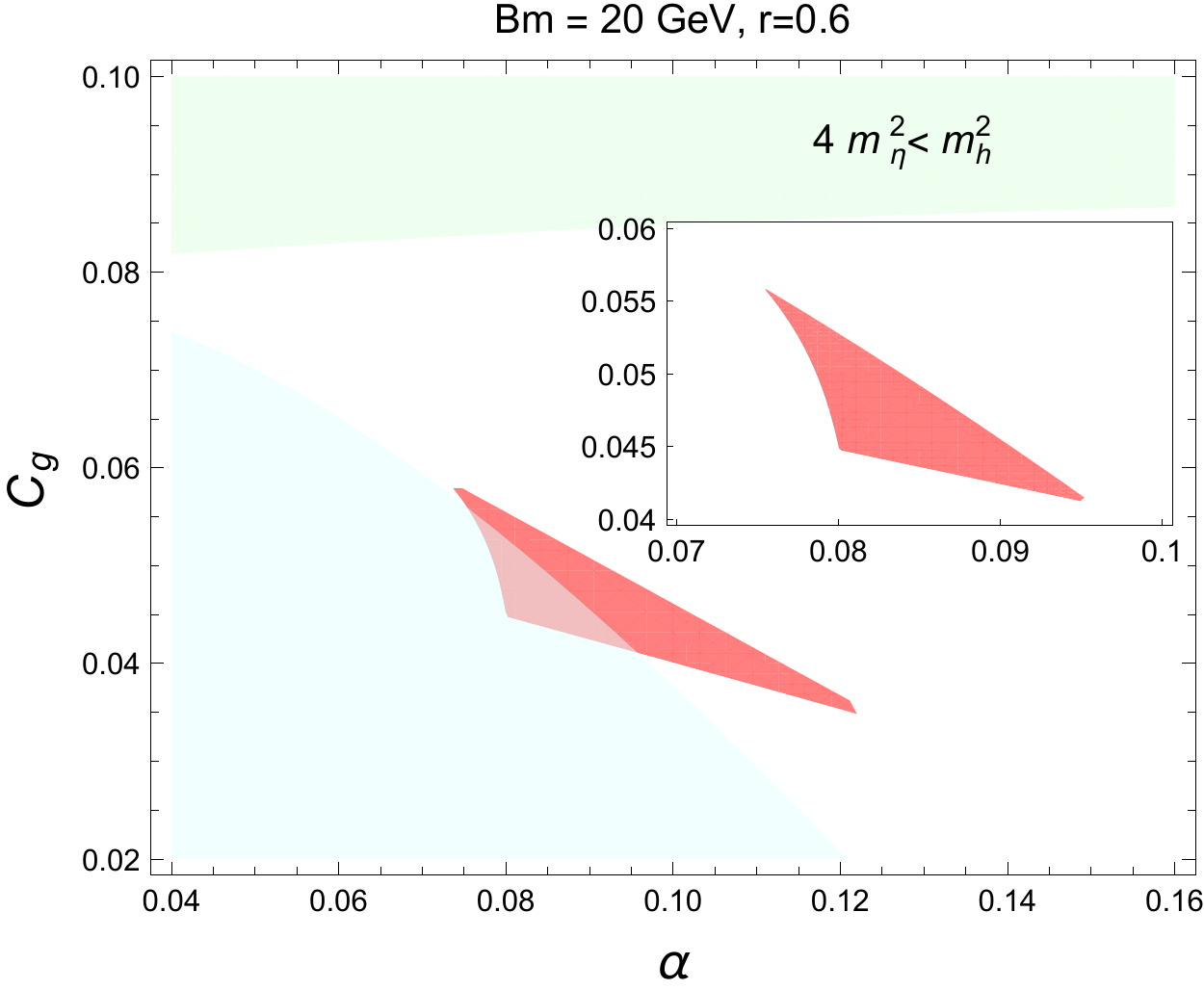}	
		\qquad \includegraphics[height=5.8 cm, 
		width=7.0cm]{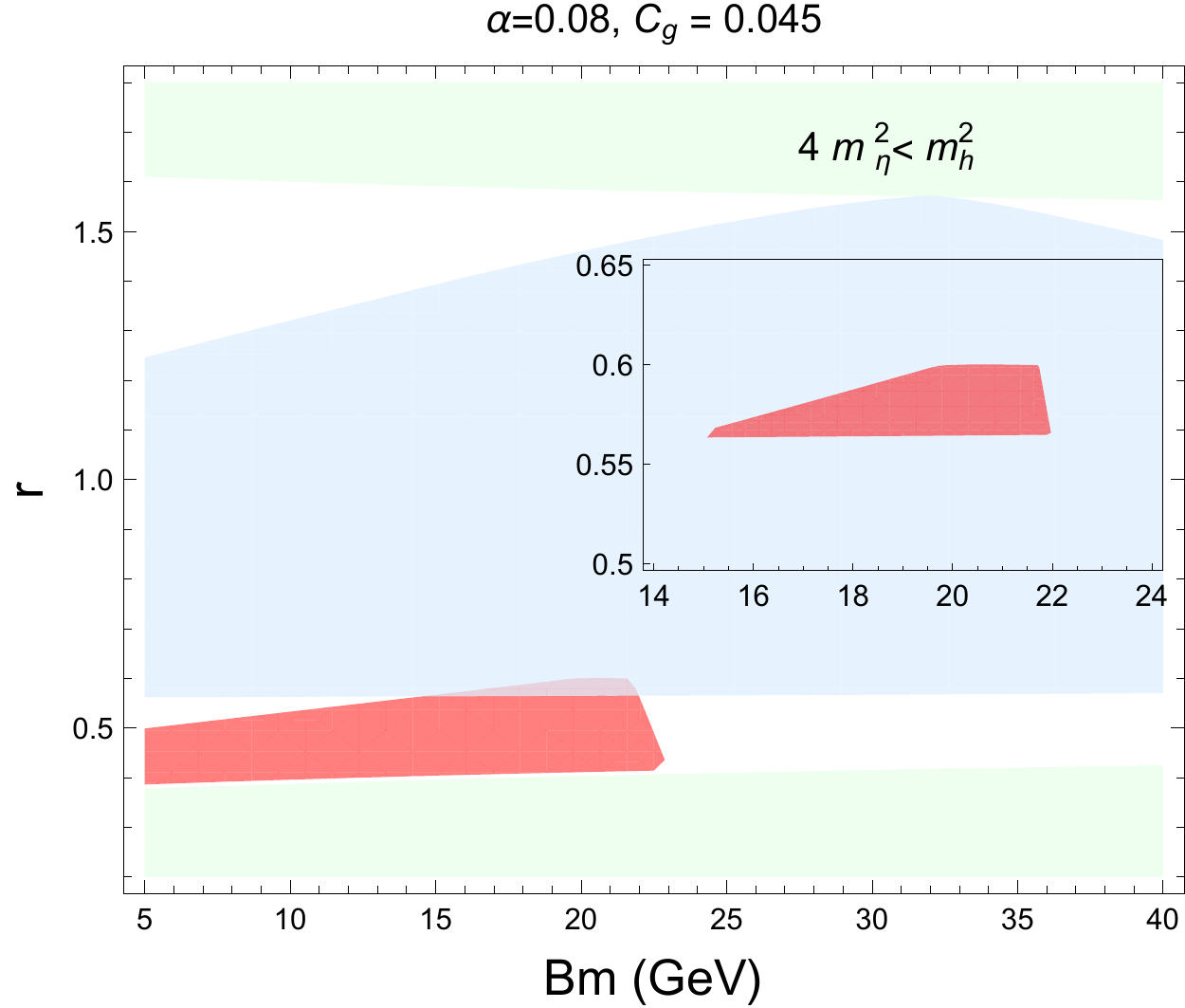}
		\caption{The red wedge in these plots indicates  the region  satisfying  the  relic density bound $\Omega h^2 < 0.123$,   along with non-tachyon  and negligible co-annihilation conditions imposed. The cyan bands are permitted by the Direct Detection bound while the green bands are the regions with $2 m_\eta < m_h$ ($m_\eta^2 <0$ included). The inset plots are the zoom-in of  regions  that satisfy all the constraints. } \label{fig: relic}
\end{figure*}

Due to the DM parity described in Eq~\eqref{eq: parity}, all odd particles  $\eta_2$ and $\phi_i = (A_0,  H^0, H^{\pm}) \in H_2$  participate in the thermal equilibrium before the freeze out.  There exist a vertex of $\eta_2$-$A_0$-$\lambda_0$, thus via the Yukawa operators for the light quarks,  the heavier component will quickly decay into the lightest mass eigenstate after freeze-out.  In principal, the effective averaged cross-section need to  take into account all  co-annihilation processes~\cite{Griest:1990kh, Edsjo:1997bg}.

However,  the mass hierarchy in the spectrum is of paramount importance for co-annihilation.
In our model,  the inert Higgs doublet  observes a mass hierarchy $m_{H_0} > m_{H^\pm} \gtrsim m_{A_0}$~\cite{Cacciapaglia:2019ixa}. In this work, we only investigate a simplified scenario $m_{A_0} > m_{\eta_2}$ with a large mass gap of $\Delta =  (m_{A_0}- m_{\eta_2})/m_{\eta_2} > 0.2$, so that the  actual DM is the singlet.  At a typical freeze out temperature $x_f \sim 25$, the ratio of number densities, $n_{\phi_i}^{\rm eq}/ n_{\eta_2}^{\rm eq}$, is  highly suppressed by  a  Boltzmann factor  $(1+ \Delta )^{3/2} \exp(-\Delta \, x_f) \sim 1/200$. Since the direct annihilation $ \langle \sigma_{\eta_2 \eta_2} v \rangle $  is not subdominant, the co-annihilation effect can be safely neglected  and one can only consider the direct annihilation of the singlet $\eta_2$. 
The dominant channels are:
\begin{align}
 \langle \sigma_{\rm eff} v \rangle & \simeq    \langle \sigma v (\eta_2 \eta_2 \to V V)  \rangle  +   \langle \sigma v (\eta_2 \eta_2 \to h h  )  \rangle 
 + \nonumber \\
&  \langle \sigma v (\eta_2 \eta_2 \to \bar{f} f )  \rangle   +   \langle \sigma v (\eta_2 \eta_2 \to \eta_{p/m} \eta_{p/m} )  \rangle\,, 
  \end{align}
where $\eta_{p/m}$ are linear combinations of the $\mathbb{Z}_2$--even $\eta_{1,3}$, which we should include in because they are typically much lighter than other pNGBs and will eventually decay into SM particles~\cite{Cacciapaglia:2019ixa}. 
The invariant cross-section  and the M\o ller velocity in the lab frame  are:
  \begin{align} 
 \sigma(\eta_2 \eta_2 \to X Y)  = \frac{1}{64 \pi^2 s^{3/2} \sqrt{s- 4 m_\eta^2} } \int d\Omega |{\cal{\bar M}}_{X Y}|^2  \times \nonumber \\ 
  \left[s- (m_{X}+m_{Y})^2 \right]^{1/2} \left[s- (m_{X}-m_{Y})^2\right]^{1/2} \,, 
 \end{align}
and
 \begin{equation}
v_{\rm lab}  =   \sqrt{(p_a \cdot p_b)^2  - m_a^2 m_b^2}/ (E_a E_b)\,.
  \end{equation}
  The amplitudes $|{\cal{\bar M}}_{X Y}|^2$ are given in the Appendix~\ref{app: amplitude}, as function of all  relevant couplings in the model.
The key ingredient for  relic density  is the thermal averaged cross-section, which can be evaluated by an integral (without velocity expansion of  $\sigma v_{\rm lab}$)~\cite{Gondolo:1990dk}:
\begin{align} 
& \langle\sigma  v_{\rm lab}\rangle = \frac{1}{8 m_\eta^4 T K_2(x)^2} \int_{4m^2_\eta}^\infty  ds ~ \sigma \sqrt{s} (s-4 m_\eta^2)  K_1\left(\frac{\sqrt s}{T}\right) \label{eq:thermal}
\end{align}
with $K_{1,2}(x)$ being the modified Bessel functions of the second kind.  In the region far from the resonance (Higgs mass), like the singlet DM with a mass of hundreds of GeV,  the integral approach precisely matches with the $s$, $p$ and $d$-wave  limits~\cite{Srednicki:1988ce}.
 
We now discuss the impact of Direct Detection experiments, which are sensitive to the recoil energy deposited by the scattering of DM to nucleus.
In our model, the relevant interactions at the microscopic level  are:
\begin{equation}
\mathcal{L} \supset - \lambda_{h \eta_2^2} h \eta_2^2    - \lambda_{h \bar{q} q} h \bar{q} q   +  \frac{C_q  m_q}{f^2} \eta^2_2 \bar{q} q\,.  
\end{equation} 
As already mentioned in the introduction, we work in the basis where derivative couplings of the DM candidate to a single Higgs boson are absent, thus all the above couplings explicitly break the shift symmetry associated with the Goldstone nature of the stable scalar.
Besides the Higgs portal, we also have direct couplings to quarks  from non-linearities in the pNGB couplings (for light quarks,  the last term comes from an effective Yukawa coupling). Thus, the spin-independent DM-nucleon cross section (factoring out  $\frac{\mu_T^2}{\mu_N^2} A^2$ with respect to  $\sigma^{\rm SI}_{\rm Nucleus}$) can be parameterised as~\cite{Han:1997wn, Frigerio:2012uc}:
\begin{eqnarray}
\sigma_{\rm SI} =  \frac{m_N^2}{\pi m_{\eta_2}^2} \left( \frac{m_N m_{\eta_2}}{m_N + m_{\eta_2}} \right)^2  \frac{(Z f_p + (A-Z) f_n  )^2}{A^2}\,,
\end{eqnarray}
where  $f_{p,n}$  characterise the interaction between $\eta_2$ and nucleons: 
\begin{eqnarray}
f_{p,n} = \sum \limits_{q} f^{(p,n)}_q \left[ \frac{C_q}{f^2} + \frac{\lambda_{h \eta_2^2}}{v m_h^2} \right] 
\end{eqnarray}
with $q$ summing over all quark flavours.  Note that setting $C_q=0$ will exactly match the result in~\cite{Cline:2013gha}. For the form factors  $f_q^{(p,n)} = \frac{m_q}{m_N} \langle N | \bar q q | N \rangle$~\cite{Drees:1993bu}, we will use the values from \cite{Cheng:2012qr} for the light quarks, $u$, $d$ and $s$, while for heavy flavours the form factor can be computed via an effective coupling to gluons at one loop, giving
\begin{equation}
f_{c/b/t}^{(p,n)} = \frac{2}{27} f_{TG}^{(p,n)}= \frac{2}{27} \left( 1- \sum \limits_{q = u,d,s} f_q^{(p,n)}\right)\,.
\end{equation}

For this model, the $\eta_2$ annihilation will mainly proceed in the following channels  $\eta_2 \eta_2 \to W^+ W^-$, $ ZZ$, $ h h $, $\bar{t} t$, $\bar b b$,  and $\eta_{p,m} \eta_{p,m}$.   The interactions and masses of pNGBs are determined by four parameters: $(\alpha, C_g, Bm,  r)$, where we have neglected $\gamma$ and set $\delta=0$. The latter indicates a universal mass for all underlying fermions, and the $\delta$ only affects the mass splitting between $\eta_p$ and $\eta_m$ with minor impact on this analysis. Note that the  $r$ will influence  the mass spectrum of DM parity odd particles $\eta_2, A_0, H_{0, \pm}$ but not others~\cite{Cacciapaglia:2019ixa}.  These parameters are preliminarily constrained by the non-tachyon conditions, i.e. $m_{\pi}^2 >0 $ for all pNGBs.  

In Figure~\ref{fig: relic}, we show the prospect for $\eta_2$ to play the role of  DM, as opposed to the current  bounds from direct detection. There is no constraint from the Higgs  decay width since all singlet masses are heavier than $m_h$.  For the two panels,  we highlight the regions in red that satisfy the relic density bound $\Omega h^2 < 0.123$, as well as non-tachyon conditions and $m_{A_0}>1.2 \, m_{\eta_2}$.  While the light blue regions are permitted by the direct detection since we impose $\sigma_{\rm SI} < \sigma_{\rm SI}^{\rm exp}$.  Only the overlaps between the red and blue  are viable, where the relic density  almost saturates $\Omega_{\eta_2} h^2 \sim 0.12 $,  thus it is not necessary to rescale  $\sigma_{\rm SI}$ by a  factor of $\Omega_{\eta_2}/\Omega_{\rm DM}$.  The green regions indicate near the edge   $2 m_{\eta_2} = m_h$,  the Higgs resonance is active, but  they are far from the viable region, thus validate our  analytical computation of the relic density. The region allowed by all the bounds, therefore, is fairly limited.  Without fixing the parameters of $(C_g, Bm)$, we get two branches  centered around  $(\alpha, r) \approx (0.08 \pm 0.01, 0.6 \pm 0.15)$ or $(0.02 \pm 0.01, 0.35 \pm 0.05)$, which implies  $f \approx v/\alpha \sim 3.0$ or $12.0$~TeV. We want to stress  that the DM mass  can be computed for each point of $(\alpha, C_g, Bm,  r)$, so it varies over the slices of parameter space  in the figure.

\begin{figure}[tb]
	\centering 
		\includegraphics[height=5.9 cm, width=7.0cm]{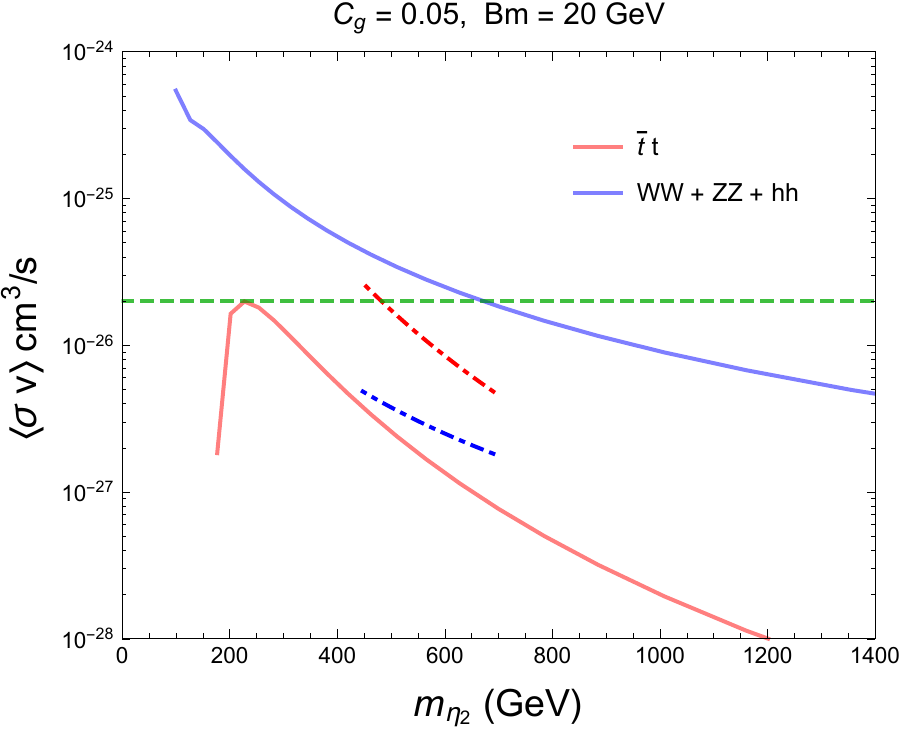}	
	\caption{Averaged cross-section in the channels $\bar{t}t$ (red) and combined $W^+ W^-, ZZ, hh$ (blue) for two benchmark points with $r= 0.4$ (dashed) and $r= 0.6$ (solid) as a function of $m_{\eta_2}$. The dashed green line is  the reference  value $\langle \sigma v   \rangle =  2.0 \times 10^{-26} \, cm^3/s $ for a relic density $\Omega h^2 \sim  0.12$.} \label{fig: mass1}
\end{figure}

\begin{figure}[tb]
	\centering 
	\includegraphics[height=5.6 cm, width=7.0cm]{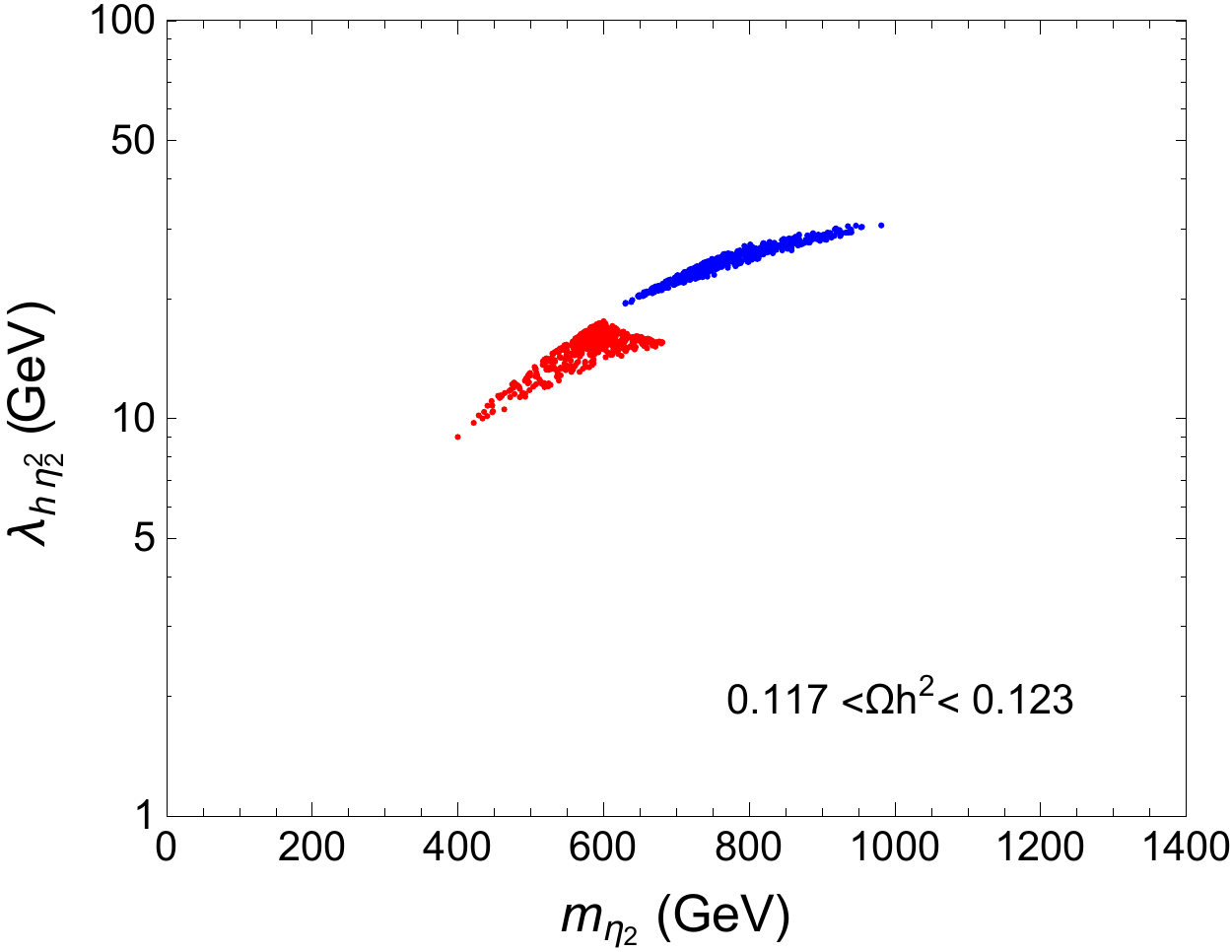}
	\caption{Correlation between the Higgs portal coupling  $\lambda_{h \eta^2}$ and  the DM mass $m_{\eta_2}$  after imposing all the necessary conditions, i.e. $0.117<\Omega h^2 < 0.123$, non-tachyon,  negligible co-annihilation as well as direct detection bound. The red dots represent  scenarios where the $\langle \sigma ({\bar{t} t}) v \rangle $  is dominant, while the blue dots stand for cases where the combined  channels $W^+W^-, ZZ, hh$  contribute the most. } \label{fig: mass2}
\end{figure}

To better investigate the DM physics, we selected two illustrative benchmark points in Figure~\ref{fig: mass1}: we show  the thermal averaged cross-section  $\langle \sigma v \rangle$ for the $\bar t t $ channel and for the combined $W^+W^-, ZZ, hh$ channels at $r = 0.4$ and $r=0.6$ respectively, with $C_g =0.05$ and $Bm = 20 \mbox{GeV}$  fixed.  Note that the intersection with the dashed green line indicates points saturating the observed relic density. We do not add the direct detection bound for Figure~\ref{fig: mass1}. For  $r =0.4$, the lower limit of $m_{\eta_2}$ derives from the non-tachyon condition, $m_{\eta_m}^2>0$, while the upper limit is bounded by the requirement of  $m_{A_0} > 1.2\ m_{\eta_2}$.  At this benchmark, the $\bar t t$ cross-section is dominant, reaching above $80\%$ percent of the total contribution. The dot-dashed red line intersects with the reference cross-section line for a DM mass of $ m_{\eta_2} \sim 500$ GeV. Instead, for $r = 0.6$, the averaged cross-section mainly comes from the combined di-bosons and di-Higgs channels, with the blue solid line intersecting the reference value for a DM mass around $m_{\eta_2} \sim 700$~GeV.  As inferred from Figure~\ref{fig: relic},  a smaller mass with a larger cross section  is excluded by the direct detection. What this plot reveals is that due to the  model-specific couplings, the DM cross section is dominated by the $\bar t t$ channel for lower masses, and by the di-boson one for larger masses. For a further illustration, in Figure~\ref{fig: mass2}, we show the results  of a random scan of  $(\alpha, C_g, Bm, r)$, where we retain only the points passing all the constraints, i.e. relic density, direct detection and no-tachyon conditions.
As expected, we find two distinct branches of viable points:  for $400 < m_{\eta_2} < 600$ GeV, the $\bar t t$ channel dominates, while for $600 < m_{\eta_2}< 1000$~GeV, it is the di-boson $W^+W^-, ZZ, hh$ that dominates the annihilation cross section. Note that this latter section is similar to the traditional Higgs portal singlet model. One new ingredient in our model is the annihilation into the lighter even pNGBs $\eta_p$ and $\eta_m$: we have checked that this channel  is always subleading, with a contribution of less that $3\%$ for all points. Also, since the quartic couplings are relatively small, the cubic coupling $\lambda_{h \eta_2^2}$  is always close to  the upper limit of the direct detection bound in order to put the relic density into the correct ballpark. This implies that the singlet DM candidate in our model is on the verge of being excluded  by future direct detection experiments, like XENONnT~\cite{Aprile:2018dbl} and LUX-ZEPLIN~\cite{Akerib:2018lyp}. The projected sensitivity of the latter will improve the reach of the current XENON1T experiment by at least one order of magnitude. We found out that in such a situation there will be no overlap remaining in Fig.~\ref{fig: relic}, and this DM scenario will be excluded. However, the same might not happen to the other DM candidate, i.e. $A_0$ in the inverted scenario of $m_{A_0} < m_{\eta_2}$. We leave this case for further investigation.

We should  also consider the indirect detection signals, since  the gamma ray line spectrum from the Galactic center observed by Fermi-LAT and H.E.S.S.  is possible to impose constraints for  a DM mass in the interval of  $(10, 10^3)$ GeV.  According to the recent measurements, provided $400 < m_{\rm DM}< 1000$ GeV,  the lowest upper limits  on the DM annihilation cross section $\langle \sigma v \rangle $ for the $t \bar t$, $W^+ W^-$, $ZZ$ and $hh$  channels are around $ 10^{-25}  ~\mbox{cm}^3/s$ at $m_{\rm DM} \simeq 400$ GeV from Fermi-LAT~\cite{Ackermann:2015zua} and $ 4.0 \times 10^{-26}  \mbox{cm}^3/s$ at $m_{\rm DM} \simeq 1$ TeV from H.E.S.S~\cite{Abdallah:2016ygi}. In fact, the upper bounds in the mass region of interest $m_{\rm DM } \sim (400, 10^3)$ GeV, given by both telescopes from those concerning final states, are all well above  the reference value $2.0 \times 10^{-26} \mbox{cm}^3/s$. This statement is  consistent with the earlier analyses~\cite{Abazajian:2010sq, Abazajian:2010zb}, where the relevant bounds are one order of magnitude larger.  Thus in our model with the viable region almost saturating the relic density, the indirect detection can barely exclude any interesting parameter space.  We would like to briefly comment on  the collider phenomenology. Although the DM mass range  is potentially accessible at colliders, including the LHC, the specific searches aiming at the DM pair production plus one additional SM particle are not available. As the masses being in the multi hundred GeV ballpark, the pair production cross-sections are very small and may only be accessible to the high luminosity run or at future high-energy colliders. The model  also contains a light $\mathbb{Z}_2$--even pseudoscalar, whose mass can go down to one GeV. The collider phenomenology of this state has been discussed in Ref.~\cite{Cacciapaglia:2019ixa}.

\section{Conclusion} \label{conclusion}

In this work, we analysed the properties of a pseudo-scalar singlet  Dark Matter candidate that emerges as a pseudo-Nambu-Goldstone boson from a composite Higgs model, based on the coset $SU(6)/SO(6)$.  Since the light quarks, in particular the bottom,  have to obtain their masses via an effective Yukawa operator originating from four fermion interaction, the custodial symmetry is unavoidably broken.  We have proved that  all the tadpole terms in a generic vacuum with  $(\alpha, \gamma)$ angles  will vanish  after imposing the minimum conditions.  More importantly,  the $\gamma$ value is   suppressed by an order of $m_b^2/m_h^2$, making the custodial symmetry breaking  well under control. Our result should also apply to the CHM model in $SU(5)/SO(5)$,  which shares the same bi-triplet structure but without DM candidate.

Like most Higgs portal singlet models, the parameter space in this model is tightly constrained, mainly by Direct Detection experiments. Yet, due to the non-linear nature of the DM couplings to the SM particles, we find that a small region of the parameter space is still allowed, with masses $400 < m_{\rm DM} < 1000$~GeV. For masses below $\sim 600$~GeV, the annihilation cross section is dominated by $\bar t t$, unlike the traditional models, while larger masses go back to dominant di-boson channels.  This pattern comes from  the  complicated pNGB  couplings. Furthermore, requiring the correct relic density leaves the allowed regions within  reach of future experiments, XENONnT and LUX-ZEPLIN, which might be able to exclude the $\eta_2$ as a DM.

Finally, the $SU(6)/SO(6)$ model has another DM candidate in an inert second Higgs doublet, if it is lighter than the singlet. The phenomenology of this state is similar to the traditional inert Higgs doublet model~\cite{Belyaev:2016lok, LopezHonorez:2006gr}, which is also tightly constrained by observations. We leave an exploration of this limit for future work.

\section*{Acknowledgements}
G.C.  acknowledges partial support from the Labex Lyon Institute of the Origins - LIO. The research of H.C. is supported by the Ministry of Science, ICT \& Future Planning of Korea, the Pohang City Government, and the Gyeongsangbuk-do Provincial Government. H.C also acknowledges the support of TDLI  during  the revision of manuscript.


\bibliography{su6DM}

\newpage

\onecolumngrid

\input{appendix}

\end{document}

%% file: appendix.tex
\appendix
\section{Tadpole terms in a general vacuum}\label{app: tadpole}

To ensure the DM stability,  only two CP-even neutral pNGBs in this model, i.e.  $h$ and $\lambda_0 = -\frac{i}{\sqrt{2}}\left(\Lambda_0 - \Lambda_0^* \right)$ can  develop VEVs. This means that we  set  $\beta =0$  so that the second Higgs doublet remains inert.  The pNGB matrix is parametrized as:
\begin{equation}
\Sigma_{\alpha, \gamma} (x) = U (\alpha, \gamma) \cdot \Sigma(x) \cdot U^T (\alpha, \gamma)= U (\alpha, \gamma) \cdot e^{i\frac{2 \sqrt{2}}{f} \Pi (x)} \cdot \Sigma_{\rm EW} \cdot U^T (\alpha, \gamma)\,, \label{eq: sigma}
\end{equation}
where the  $\Sigma_{\rm EW}$ is rotated away  to be misaligned with the direction of $SU(2)_L \times U(1)_Y$ by $U(\alpha, \gamma)$. In analogy to the $\beta$ rotation, the $SU(6)$ transformation can be decomposed into:
\begin{eqnarray}
 U(\alpha, \gamma) = U_{\gamma}\cdot U_{\alpha} \cdot U_{\gamma}^\dag = e^{i \gamma \sqrt{2} S_{8}} \cdot e^{i \alpha \sqrt{2} X_{10}}\cdot e^{-i \gamma \sqrt{2} S_{8}}  
 \end{eqnarray}
with  $U_{\gamma}= e^{i \gamma \sqrt{2} S_{8}}$   defined in term of an unbroken generator.  The inner  $U_\gamma^\dag$ operation on $\Sigma(x)$ can be fully absorbed by the pNGB field redefinition. Thus it is  the outer $U_{\gamma}$  that determines the $\gamma$ dependence. And Eq.(\ref{eq: sigma}) can be re-written as:
\begin{equation}
\Sigma_{\alpha, \gamma} (x) = U_{\gamma} \cdot U_{\alpha} \cdot  \tilde{\Sigma}(x) \cdot U_{\alpha}^T \cdot U_{\gamma}^T = U_{\gamma} \cdot U_{\alpha} \cdot e^{i\frac{2 \sqrt{2}}{f} \tilde{\Pi} (x)} \cdot \Sigma_{\rm EW} \cdot U_{\alpha}^T \cdot U_{\gamma}^T \,.
\end{equation}
with $\tilde{\Sigma} (x)  =  U_{\gamma}^\dag \cdot \Sigma (x) \cdot U_{\gamma}^*$.  Thus  we can find out  an exact mapping for $\Pi(x) \to \tilde{\Pi}(x)$ in terms of pNGB fields. The field redefinition can be split into several blocks and leaving the pion fields $H_{\pm, 0}$ unchanged.  For $h$ and  $\lambda_0$, they transform in a $SO(2)$ rotation defined by  $\gamma$: 
\begin{eqnarray}
\tilde{h} &=& h \cos (\gamma) - \lambda_0 \sin (\gamma)  \nonumber \\
\tilde{\lambda}_0 &=&  h \sin (\gamma) + \lambda_0 \cos (\gamma ) 
\end{eqnarray}
This also applies to the DM candidates $A_0$ and $\eta_2$:
\begin{eqnarray}
\tilde{A}_0 &=& A_0 \cos (\gamma) -   \eta_2 \sin (\gamma) \nonumber \\
\tilde{\eta}_2 &=& \eta_2 \cos(\gamma)  + A_0 \sin (\gamma) 
\end{eqnarray}
The charged Goldstone $G_+$ eaten by $W^+$ can mix with the $\Lambda_+$ and $\phi_+$ in the bi-triplet.
\begin{eqnarray}
\tilde{\Lambda}_+ &=& \Lambda_{+} \cos ^2\left(\frac{\gamma }{2}\right)- \varphi_{+} \sin^2\left(\frac{\gamma }{2}\right)-\frac{i }{\sqrt{2}} G_{+} \sin (\gamma )
\nonumber \\
\tilde{\varphi}_+ &=&  \varphi_{+} \cos^2\left(\frac{\gamma }{2}\right)- \Lambda_{+} \sin ^2\left(\frac{\gamma }{2}\right)-\frac{i }{\sqrt{2}} G_{+} \sin (\gamma )
\nonumber \\
\tilde{G}_+ &=& G_{+} \cos (\gamma )-\frac{i }{\sqrt{2}} (\Lambda_{+}+ \varphi_{+}) \sin (\gamma) 
\end{eqnarray}
Finally the neutral Goldstone $G_0$  mixes with $\eta_{1,3}$, $\lambda$ and $\varphi_0$ under the $\gamma$ rotation. With the definition $\eta_G \equiv \frac{1}{2 \sqrt{2}}  \left(\sqrt{3} \eta_{1}- \sqrt{2} \eta_{3} + \sqrt{2} \lambda - \varphi_{0}\right)$, we can obtain:
\begin{eqnarray}
\tilde{G}_0 &= & G_{0} \cos (2 \gamma ) - \eta_G \sin (2 \gamma ) \nonumber \\
\tilde{\eta}_1 &=& \eta _1- \frac{1}{2} \sqrt{\frac{3}{2}} \left( 2 \eta_G \sin ^2(\gamma ) - G_{0} \sin (2 \gamma ) \right)  \nonumber \\
\tilde{\eta}_3 &=& \eta _3 + \frac{1}{2} \left( 2 \eta_G   \sin^2(\gamma ) -  G_{0} \sin (2 \gamma ) \right) \nonumber \\
 \tilde{\lambda} &=&\lambda -   \frac{1}{2} \left(2 \eta_G \sin ^2(\gamma ) -  G_{0} \sin (2 \gamma )\right) \nonumber \\
\tilde{\varphi}_0 &=& \varphi _0 + \frac{1}{2 \sqrt{2} } \left( 2  \eta_G  \sin ^2(\gamma ) -  G_{0}
   \sin (2 \gamma )\right)  \label{eq: neutral}
\end{eqnarray}
Note that the last four expressions in Eq.(\ref{eq: neutral}) give: $\tilde{\eta}_G = \eta_G \cos (2 \gamma) + G_0 \sin (2 \gamma)$, explictly orthogonal to $\tilde{G}_0$.  It turns out to be  easier to calculate the  potentials in  the new basis of $\tilde{\Pi}(x)$. We can demonstrate that for each type of potential in  a generic vacuum with $(\alpha, \gamma)$ angles,  the coefficient  of $\tilde{h}$ tadpole term is equal to $ \frac{1}{f} \frac{\partial V_0(\alpha, \gamma)}{\partial \alpha}$, while the coefficient of  $\tilde{\lambda}_0$ tadpole term  is equal to $- \frac{1}{v} \frac{\partial V_0(\alpha, \gamma)}{\partial \gamma}$. 

\vspace{10 pt}

\noindent {\bf The gauge potential}:
\begin{eqnarray}
V_{g} &=& \frac{C_{g}}{4}f^{4}\big(g_2^{2}Tr[T_{L}^{i}\Sigma(T_{L}^{i}\Sigma)^{*}]+g_1^{2}Tr[Y\Sigma(Y\Sigma)^{*}]\big)
\end{eqnarray}
Expand the pion matrix,  the vacuum term at the lowest order is 
\begin{eqnarray}
\mathcal{V}_{g,0} (\alpha , \gamma)&=&  - \frac{ C_{g}f^{4}}{32} \left(4 \sin ^2(\alpha ) \cos (2 \gamma ) \left(2 \cos (2 \alpha )
   \left(g_{1}^2+ g_{2}^2\right)+ g_{1}^2+3 g_{2}^2\right)  + 2 \cos (2 \alpha ) \left(g_{1}^2+7 g_{2}^2\right) \right. \nonumber \\ &+& \left. 2 \cos (4 \alpha )
   \left(g_{1}^2+ g_{2}^2\right) + 4 \left(g_{1}^2+2 g_{2}^2\right)\right) 
\end{eqnarray}
The tadpole terms are obtained by expanding till the linear order:
\begin{eqnarray}
V_{g} (\tilde{h}) &=& \frac{C_g f^3}{8}  \bigg(4 \sin (4 \alpha ) \sin ^2(\gamma )
   \left(g_{1}^2+g_{2}^2\right)+\sin (2 \alpha ) \left( g_{1}^2+7 g_{2}^2 +\cos (2 \gamma ) (g_{1}^2- g_{2}^2) \right)\bigg) \,  \tilde{h} \nonumber \\
   & = & \frac{1}{f } \frac{\partial \mathcal{V}_{g, 0}(\alpha, \gamma)}{\partial \alpha} \, \tilde{h}
\\
V_{g} ( \tilde{\lambda}_0) &=& - \frac{C_g}{4} f^3 \sin (\alpha ) \sin (2 \gamma ) \left(2 \cos (2 \alpha )
   \left(g_{1}^2+ g_{2}^2\right)+g_{1}^2+3 g_{2}^2\right) \, \tilde{\lambda}_0 \nonumber \\
&= & - \frac{1}{f \sin \alpha} \frac{\partial \mathcal{V}_{g, 0} (\alpha, \gamma)}{\partial \gamma} \, \tilde{\lambda}_0 = -\frac{1}{v} \frac{\partial \mathcal{V}_{g, 0}( \alpha,  \gamma)}{\partial \gamma} \, \tilde{\lambda}_0
\end{eqnarray}

\vspace{10 pt}

\noindent {\bf The bottom Yukawa potential}:
\begin{eqnarray}
V_{b} &=& C_b  f^{4} \sum_{\delta}\bigg|Y_{b1}Tr[P_{b1}^{\delta}.\Sigma(x)]+Y_{b2}Tr[P_{b2}^{\delta}.\Sigma(x)]\bigg|^{2} \,,
\end{eqnarray}
The vacuum term is:
\begin{eqnarray}
\mathcal{V}_{b, 0} (\alpha , \gamma)&=&   2  C_b f^4  Y_{\text{b1}} \left(Y_{\text{b1}}\right){}^*   \sin ^2(\alpha )  \cos ^2(\gamma ) (\cos (\alpha )-\sin (\alpha ) \sin (\gamma ))^2
\end{eqnarray}
The tadpoles terms are:
\begin{eqnarray}
V_{b} (\tilde{h}) &=&  4 C_b f^3 Y_{\text{b1}} \left(Y_{\text{b1}}\right){}^*   \sin (\alpha )  \cos
   ^2(\gamma ) (\cos (\alpha )-\sin (\alpha ) \sin (\gamma )) (\cos (2 \alpha )-\sin (2
   \alpha ) \sin (\gamma ))  \, \tilde{h} \nonumber \\
   & = & \frac{1}{f } \frac{\partial \mathcal{V}_{b, 0}(\alpha, \gamma)}{\partial \alpha} \, \tilde{h}
\\
V_{b} (\tilde{\lambda}_0) &=&  4 C_b f^3 Y_{\text{b1}} \left(Y_{\text{b1}}\right){}^* \sin (\alpha)   \cos(\gamma) (\sin (\alpha ) \cos (2 \gamma )+\cos (\alpha ) \sin (\gamma )) (\cos (\alpha )-\sin(\alpha ) \sin (\gamma ))  \, \tilde{\lambda}_0 \nonumber \\ 
   &=&  -\frac{1}{v} \frac{\partial \mathcal{V}_{ b, 0}( \alpha,  \gamma)}{\partial \gamma} \, \tilde{\lambda}_0
\end{eqnarray}

\vspace{10 pt}

\noindent {\bf The top spurion potential}:
\begin{eqnarray}
V_{t} = \frac{C_{LL} f^4 }{4}\,\,  Tr[\bar{D}_{L}^{T}\cdot \Sigma^{\dagger}\cdot D_{L}\cdot \Sigma] + \frac{C_{RR} f^4 }{4}\,\, Tr[\bar{D}_{R}^{T}\cdot \Sigma^{\dagger}\cdot D_{R}\cdot \Sigma] \,.
\end{eqnarray}
Setting $C_{LL} = C_{RR} =1$, the vacuum term is
\begin{eqnarray}
\mathcal{V}_{t, 0} &=& \frac{f^4}{16} \left(2 \sin ^2(\alpha ) (2 \cos (2 \alpha )+3) \cos (2 \gamma )-5 \cos (2
   \alpha )-\cos (4 \alpha )-2\right)  \left | Q_{A} \right |{}^2  \nonumber \\
   &+& \frac{f^4}{256}  \left(8 \cos (4 \alpha ) \cos ^4(\gamma )+8 \cos (2
   \alpha ) \sin ^2(2 \gamma )-4 \cos (2 \gamma )+3 \cos (4 \gamma )\right)     \left | R_{S} \right |{}^2 
  \nonumber \\  &+&  \rm{Constant Term}
\end{eqnarray}
The tadpole terms are:
\begin{eqnarray}
V_{t} (\tilde{h}) &=&  \frac{ f^3 }{16} \sin (\alpha ) \cos (\alpha ) \big(4 \left(8 \cos (2 \alpha ) \cos
   ^2(\gamma )+\cos (2 \gamma )+5\right) \left | Q_{A} \right | {}^2 \nonumber \\ &+& 
   \left | R_{S} \right |{}^2 \left(-8 \cos (2 \alpha ) \cos ^4(\gamma )+\cos (4 \gamma
   )-1\right)\big) \, \tilde{h}  \nonumber \\
   &=&  \frac{1}{f } \frac{\partial \mathcal{V}_{t, 0}(\alpha, \gamma)}{\partial \alpha} \, \tilde{h}
\end{eqnarray}

\begin{eqnarray}
V_{t} (\tilde{\lambda}_0) &=& \frac{ f^3}{4} \sin (\alpha ) \sin (2 \gamma ) \big((2 \cos (2 \alpha )+3)
   \left | Q_{A} \right | {}^2 \nonumber \\ &+& \left | R_{S} \right | {}^2 \left(\sin ^2(\alpha ) \cos (2 \gamma )-\cos ^2(\alpha )\right)\big) \, \tilde{\lambda}_0
   \nonumber \\
   &=&   -\frac{1}{v} \frac{\partial \mathcal{V}_{t, 0}( \alpha,  \gamma)}{\partial \gamma} \, \tilde{\lambda}_0
\end{eqnarray}

\vspace{10 pt}

\noindent {\bf The  mass term potential}: 
\begin{eqnarray}
V_{m} = - \frac{B f^3}{2 \sqrt{2}} Tr[M^\dagger \cdot \Sigma] + \mbox{h.c.} 
\end{eqnarray}
with 
\begin{eqnarray}
M  =\left(
\begin{array}{cc|cc}
& i m_{2}\sigma_2 &  \\
-i m_{2} \sigma_2 &  & \\ \hline
& & m_{1}\mathbbm{1}_{2} 
\end{array} \right)\,,
\end{eqnarray}
for  $m_1 = m_2$,  the mass matrix $M$  is aligned with $\Sigma_{\rm EW}$.   First the vacuum term in the general case is:
\begin{eqnarray}
\mathcal{V}_{m,0} &=& \frac{B f^3}{2 \sqrt{2}} \left(m_2 \left(2 \sin ^2(\alpha ) \cos (2 \gamma )+3 \cos (2 \alpha )+5\right)  - m_1 \left(2 \sin ^2(\alpha ) \cos (2 \gamma )-\cos (2 \alpha )-3\right) \right)  
\end{eqnarray}
The tadpole terms read:
\begin{eqnarray}
V_{m} (\tilde{h}) &=& \frac{B f^2}{\sqrt{2}}  \sin (2 \alpha ) \left(m_2 (\cos (2 \gamma)-3)-2 m_1 \cos ^2(\gamma )\right) \,  \tilde{h} \nonumber \\
&=&  \frac{1}{f } \frac{\partial \mathcal{V}_{m, 0}(\alpha, \gamma)}{\partial \alpha} \, \tilde{h} 
\\
V_{m} (\tilde{\lambda}_0) &=&   B f^2  \sqrt{2} \left(m_2-m_1\right) \sin (\alpha ) \sin (2 \gamma ) \, \tilde{\lambda}_0 \nonumber \\
&=& -\frac{1}{v} \frac{\partial \mathcal{V}_{m, 0}( \alpha,  \gamma)}{\partial \gamma} \, \tilde{\lambda}_0
\end{eqnarray}
 We can see for $m_1= m_2$,   there is no $\gamma$ dependence in the $\tilde{h}$ and $\tilde{\lambda}_0$ basis because $U_\gamma^T\cdot M \cdot U_\gamma = M$ holds true.  The explicit symmetry breaking is $SU(6) \to SO(6)$ and the potential is  equivalent to the one in a $\alpha$ vacuum.
 
Note that  only  for  the bottom Yukawa potential,  the tadpole term of $\tilde{\lambda}_0$ is proportional to $\cos (\gamma)$, thus  non-vanishing at $\gamma =0$; but  for the other potentials, the tadpole term of $\tilde{\lambda}_0$ is proportional to $\sin(2 \gamma)$.  Furthermore, if we change $\gamma \to - \gamma$, the minus sign for the $\tilde{\lambda}$ tadpole term will be flipped so that $V (\tilde{\lambda}_0 ) = \frac{1}{v} \frac{\partial V_0(\alpha, \gamma)}{\partial \gamma} \tilde{\lambda}_0 $.

\section{The annihilation amplitudes} \label{app: amplitude}

Here  we give all the amplitudes squared used for the relic density calculation:
\begin{eqnarray}
&& |{\cal{\bar M}}(\eta_2 \eta_2 \to W^+ W^-)|^2 = \frac{e^4 \left(12 M_{W}^4-4 M_{W}^2 s+s^2\right) }{16
   M_{W}^4 S_W^4 \left(\Gamma_{h}^2 m_{h}^2+\left(m_{h}^2-s\right)^2\right)}  \nonumber  \\ 
 & &  \phantom{xxxxxxxxxxxxx}  \times  \left( \left(2 v \cos (\alpha ) \lambda _{h\eta_2^2 }+\sin ^2(\alpha )
   \left(m_{h}^2-s\right)\right)^2+\Gamma_{h}^2 m_{h}^2 \sin ^4(\alpha )\right)
   \end{eqnarray}

\begin{eqnarray}
|{\cal{\bar M}}(\eta_2 \eta_2 \to ZZ )|^2 &=& \frac{e^4 \left(12 M_{Z}^4-4 M_{Z}^2 s+s^2\right)}{16
   C_W^4 S_W^4 M_Z^4  \left(\Gamma_{h}^2 m_{h}^2+\left(m_{h}^2-s\right)^2\right)} \nonumber \\
 &\times&   \left(\left(2 v \cos (\alpha ) \lambda _{h \eta_2^2}+\sin ^2(\alpha )
   \left(m_{h}^2-s\right)\right){}^2+\Gamma_{h}^2  m_{h}^2 \sin ^4(\alpha )\right)
   \end{eqnarray}

\begin{eqnarray}
& & |{\cal{\bar M}}(\eta_2 \eta_2 \to \bar t t )|^2  = \frac{24 \left(s-4 m_t^2\right) }{v^2 \left(\Gamma_{h}^2 m_{h}^2+\left(m_{h}^2-s\right)^2\right)} \nonumber \\
&&  \phantom{xxxxxxx} \times \left(\left( \frac{ m_{t} \cos (4 \alpha )}{\cos(2 \alpha) \cos(\alpha)}  \lambda _{h \eta_2^2}+v \left(m_{h}^2-s\right)  \lambda _{\eta_2^2 t^2}\right)^2+\Gamma_{h}^2 m_{h}^2 v^2 \lambda _{\eta_2^2 t^2}^2\right) 
\end{eqnarray}

\begin{eqnarray}
&& |{\cal{\bar M}}(\eta_2 \eta_2 \to \bar b b )|^2  = \frac{24 \left(s-4 m_b^2\right) }{v^2 \left(\Gamma_{h}^2 m_{h}^2+\left(m_{h}^2-s\right)^2\right)} \nonumber \\ && \phantom{xxxxxxxxx} \times \left( \left(  \frac{ m_{b} \cos (2 \alpha )}{ \cos(\alpha)} \lambda _{h \eta_2^2} + v   \left(m_{h}^2-s\right) \lambda _{\eta_2^2 b^2} \right)^2+\Gamma_{h}^2 m_{h}^2 v^2 \lambda _{\eta_2^2 b^2}^2\right)
\end{eqnarray}

\begin{eqnarray}
|{\cal{\bar M}}(\eta_2 \eta_2 \to h h )|^2  &=& 4 \bigg(4 \lambda _{h^2 \eta_2^2}^2  + \frac{64 \lambda _{h\eta_2^2}^4 \left(s-2 m_{h}^2\right)^2}{\left(\cos^2(\theta ) \left(4 m_{h}^2-s\right) \left(s-4 m_{\eta}^2\right) +\left(s-2 m_{h}^2\right)^2\right)^2}  \nonumber \\ & + &     \frac{12 m_{h}^2 \cos (2 \alpha ) \sec (\alpha )  \lambda _{h^2\eta_2^2} \lambda _{h \eta_2^2}\left(s- m_{h}^2\right)}{ v \left(\Gamma_{h}^2 m_{h}^2+\left(m_{h}^2-s\right)^2\right)}    +\frac{9 \lambda _{h\eta_2^2}^2 m_{h}^4 \cos ^2(2 \alpha )
   \sec ^2(\alpha )}{v^2 \left(\Gamma_{h}^2 m_{h}^2+\left(m_{h}^2-s\right)^2\right)} \nonumber \\ & -&  \frac{48 m_{h}^2 \cos (2 \alpha ) \sec (\alpha ) \lambda _{h\eta_2^2}^3 \left(2 m_{h}^4-3 m_{h}^2 s+s^2\right)}{v \left(\Gamma_{h} ^2 m_{h}^2+\left(m_{h}^2-s\right)^2\right) \left(\cos ^2(\theta )  \left(4 m_{h}^2-s\right) \left(s-4 m_{\eta}^2\right)+\left(s-2 m_{h}^2\right)^2\right)} \nonumber  \\ &+& \frac{32 \lambda _{h\eta_2^2}^2 \lambda _{h^2 \eta_2^2} \left(2 m_{h}^2-s\right)}{\cos ^2(\theta ) \left(4 m_{h}^2-s\right) \left(s-4 m_{\eta}^2\right)+\left(s-2 m_{h}^2\right)^2}  \bigg)
\end{eqnarray}

\begin{eqnarray}
|{\cal{\bar M}}(\eta_2 \eta_2 \to \eta_m \eta_m )|^2  &=& 16 \left(   \lambda _{\eta_2^2 \eta_m^2}^2+ \frac{ \left(\lambda _{h \eta_m^2}^2 \lambda _{h\eta_2^2}^2+2 \lambda _{\eta_2^2 \eta_m^2} \lambda _{h \eta_m^2} \lambda _{h\eta_2^2} \left(s-m_{h}^2\right)\right)}{\Gamma_{h}^2 m_{h}^2+\left(m_{h}^2-s\right)^2}\right) \label{eq: etam}
\end{eqnarray}

\begin{eqnarray}
|{\cal{\bar M}}(\eta_2 \eta_2 \to \eta_p \eta_m )|^2  &=& 4 \left( \lambda _{\eta_2^2 \eta_p \eta_m}^2 + \frac{\left(\lambda _{h \eta_m \eta_p}^2 \lambda _{h\eta_2^2}^2+2 \lambda _{\eta_2^2 \eta_p \eta_m} \lambda _{h \eta_p \eta_m} \lambda _{h\eta_2^2} \left(s-m_{h}^2\right)\right)}{\Gamma_{h}^2 m_{h}^2+\left(m_{h}^2-s\right)^2}  \right)
\end{eqnarray}

For  $|{\cal{\bar M}}(\eta_2 \eta_2 \to \eta_p \eta_p )|^2$, we can simply replace $\lambda_{h\eta_m^2 } \to \lambda_{h\eta_p^2}$ and $\lambda_{\eta_2^2 \eta_m^2} \to \lambda_{\eta_2^2 \eta_p^2}$ in Eq.(\ref{eq: etam}).

\section{The vertices}\label{app: vertex}
The Lagrangian relevant to DM annihilations  can be written as three parts: $\mathcal{L} = \mathcal{L}_V + {\mathcal L} _S +  \mathcal{L}_f$:
\begin{eqnarray}
\mathcal{L}_V & = &   \frac{g^2 }{8} ( 2 \,h \,v \cos (\alpha ) -  \eta_2^2  \, \sin^2 (\alpha) ) \left(2 W_\mu^- W_\mu^+ + Z_\mu^2 \sec ^2(\theta_w)\right)  
\\ \nonumber \\
{\mathcal L} _S &=& - ( \lambda_{h^2 \eta_2^2} h^2 +  \lambda_{\eta_m^2 \eta_2^2} \eta _m^2+  \lambda_{\eta_p \eta_m \eta_2^2} \eta
   _m \eta _p + \lambda_{\eta_p^2 \eta_2^2 }\eta _p^2 )  \eta _2^2  \nonumber \\
  & - &  (\lambda_{h \eta_2^2}   \eta_2^2 + \lambda_{h \eta_{p}^2} \eta_p^2   + \lambda_{h \eta_{m}^2} \eta_m^2+  + \lambda_{h \eta_{p} \eta_m} \eta_p \eta_m)  h \nonumber \\  & -&  \frac{ m_{h}^2 \cos (2 \alpha ) \sec (\alpha )}{2 v} h^3 
\\ \nonumber \\
\mathcal{L}_f & = & - \frac{4 m_t \cos (4 \alpha)}{f \sin (4 \alpha)}  h \bar{t} t - \frac{2 m_b \cos (2 \alpha)}{f \sin (2 \alpha)} h \bar b b   \nonumber \\ &+& \lambda_{\eta_2^2  \bar t t } \eta_2^2  \bar t t + \lambda_{\eta_2^2 \bar b b} \eta_2^2 \bar b b 
\end{eqnarray}
where those $\lambda$ couplings are complicated functions of $(\alpha, C_g, Bm, r)$,  imposed by the minimum $\frac{\partial V_0 (\alpha)}{ \partial \alpha }=0 $ and Higgs mass conditions after extraction from the potentials. We explicitly list their expressions as below:

\begin{eqnarray}
\lambda_{h^2 \eta_2^2} &=&   \bigg( \frac{m_{h}^2 \sec ^2(\alpha ) }{144 v^2}  \left(\left(8 r^2-50 r+20\right) \cos
   (2 \alpha )+6 r^2+7 (5-4 r) \cos (4 \alpha )-34 r+11\right)  \nonumber \\& +&
   \frac{2 \sqrt{2} B m }{9v} (2 r-1) \sin (\alpha ) (14 \cos (2 \alpha )-2 r+1) \nonumber \\ &+& \frac{2 \text{Cg} M_{w}^2 }{9 v^2} (7 (2 r-1) \cos (2 \alpha )-2 (r-1) r+1)  \left(\cos \left(2 \theta _W\right)+2\right) \sec ^2\left(\theta _W\right) \bigg)
\end{eqnarray}

\begin{eqnarray} 
\lambda_{h \eta_2^2 } &=&  \bigg( \frac{ m_{h}^2 \sec (\alpha ) }{12 v}  ((7-8 r) \cos (2 \alpha )+3 - 6 r)  \nonumber \\ & +& \frac{2 C_{g} M_{w}^2}{3 v}  \cos (\alpha ) \left(16 r -7 + (8 r -5) \cos \left(2
   \theta _W\right)\right) \sec ^2\left(\theta _W\right) \nonumber \\  &+& \frac{8}{3} \sqrt{2} B  m (2 r-1) \sin (2 \alpha )\bigg)
   \end{eqnarray}
   
   \begin{eqnarray}
\lambda _{\eta_2^2 \eta_m^2} &=&   \bigg( -\frac{m_{h}^2 \sec ^2(\alpha ) }{2880 v^2}  ((64 (7-4 r) r+636) \cos (2
   \alpha )+(128 r+41) \cos (4 \alpha )+64 (5-3 r) r+251) \nonumber \\ &+& 
   \frac{4 C_{g} M_{w}^2 }{45 v^2}  ((8 r-4) \cos (2 \alpha )-8 (r-1)
   r+39) \left(\cos \left(2 \theta _W\right)+2\right) \sec ^2\left(\theta _W\right) \nonumber \\ &+& \frac{\sqrt{2} B_{m} \sin (\alpha )}{45 v}   ((64 r+25) \cos (2 \alpha )-64 (r-1) r+159)  \bigg)
   \end{eqnarray}
   
   \begin{eqnarray}
   \lambda_{\eta_2^2 \eta_p^2} & = & \bigg( \frac{ m_{h}^2 \sec ^2(\alpha )}{30 v^2} ((r (4 r-7)+1) \cos (2 \alpha
   )+(1-2 r) \cos (4 \alpha )+r (3 r-5)+1) \nonumber \\ 
   & + & \frac{8 C_{g} M_{w}^2 }{15 v^2}  ((2 r-1) \cos (2 \alpha )-2 (r-1) r+1)
   \left(\cos \left(2 \theta _W\right)+2\right) \sec ^2\left(\theta _W\right) \nonumber \\ 
& + & \frac{2 \sqrt{2} Bm   \sin (\alpha ) }{15 v}   ((16 r-5) \cos (2 \alpha )-16 (r-1) r-9)
   \bigg)
   \end{eqnarray}
   
 \begin{eqnarray}
\lambda _{\eta_2^2 \eta_m \eta_p} &=&   \bigg( \frac{m_{h}^2 \sec ^2(\alpha ) }{120 \sqrt{6} v^2}  (-4 (4 r (4 r-7)+29) \cos
   (2 \alpha )+(32 r-31) \cos (4 \alpha )+16 (5-3 r) r-61)  \nonumber \\ 
   &+& \frac{16 \sqrt{\frac{2}{3}} C_{g} \text{Mw}^2 }{15 v^2}((1-2 r)
   \cos (2 \alpha )+2 ((r-1) r+2)) \left(\cos \left(2 \theta _W\right)+2\right) \sec
   ^2\left(\theta _W\right)  \nonumber \\ &+& \frac{8 B_{m} \sin (\alpha )  }{15 \sqrt{3} v}((15-16 r) \cos (2 \alpha
   )+16 (r-1) r+19) \bigg)
\end{eqnarray}

\begin{eqnarray}
\lambda_{h \eta_m^2} = \frac{8}{5}  \sqrt{2}  B m  \sin (2 \alpha ) 
 \quad  \quad  \lambda_{h \eta_p \eta_m} = \frac{4}{5}  \sqrt{3}  B m  \sin (2 \alpha ) 
\end{eqnarray}

\begin{eqnarray}
\lambda_{h \eta_p^2} = - \frac{4}{15}   \sqrt{2}  B m  \sin (2 \alpha )  \quad \quad    \lambda_{\eta_2^2 \bar b b} =   \frac{ m_b \sin^2 (\alpha) }{v^2} 
\end{eqnarray}

\begin{eqnarray}
 \lambda_{\eta_2^2 \bar t t} =    \frac{ m_t \sin^2 (\alpha) (2 \cos (2 \alpha )-2 r+1)}{v^2 \cos (2 \alpha ) } 
\end{eqnarray}